\documentclass[twocolumn]{raa}
\usepackage{graphicx,times}
\usepackage{natbib}
\usepackage{amssymb,amsmath}
\usepackage{multicol}
\usepackage{multirow}
\usepackage[usenames,dvipsnames]{xcolor}
\bibpunct{(}{)}{;}{a}{}{,}

\newcommand{\aref}[1]{\hyperref[#1]{Appendix~\ref*{#1}}}

\usepackage{scalerel,tikz}
\usetikzlibrary{svg.path}
\definecolor{orcidlogocol}{HTML}{A6CE39}
\tikzset{orcidlogo/.pic={
 \fill[orcidlogocol] svg{M256,128c0,70.7-57.3,128-128,128C57.3,256,0,198.7,0,128C0,57.3,57.3,0,128,0C198.7,0,256,57.3,256,128z};
 \fill[white] svg{M86.3,186.2H70.9V79.1h15.4v48.4V186.2z}
 svg{M108.9,79.1h41.6c39.6,0,57,28.3,57,53.6c0,27.5-21.5,53.6-56.8,53.6h-41.8V79.1z M124.3,172.4h24.5c34.9,0,42.9-26.5,42.9-39.7c0-21.5-13.7-39.7-43.7-39.7h-23.7V172.4z}
 svg{M88.7,56.8c0,5.5-4.5,10.1-10.1,10.1c-5.6,0-10.1-4.6-10.1-10.1c0-5.6,4.5-10.1,10.1-10.1C84.2,46.7,88.7,51.3,88.7,56.8z};
}}

\newcommand\orcidicon[1]{\href{https://orcid.org/#1}{\mbox{\scalerel*{
\begin{tikzpicture}[yscale=-1,transform shape]
\pic{orcidlogo};
\end{tikzpicture}
}{|}}}}

\usepackage[pagebackref=true]{hyperref}

\hypersetup{colorlinks = true, linkcolor = blue, anchorcolor = red, citecolor = blue, filecolor = red, pagecolor = red, urlcolor = blue}

\begin{document}

\title{Velocity Dispersion $\sigma_{\rm aper}$ Aperture Corrections as a Function of Galaxy Properties from Integral-field Stellar Kinematics of 10,000 MaNGA Galaxies}

\volnopage{ {\bf 20XX} Vol.\ {\bf X} No. {\bf XX}, 000--000}
   \setcounter{page}{1}

\author{Kai Zhu\orcidicon{0000-0002-2583-2669}\inst{1,2,3},
        Ran Li\inst{1,2,3},
        Xiaoyue Cao\orcidicon{0000-0003-4988-9296}\inst{1,2,3},
        Shengdong Lu\orcidicon{0000-0002-6726-9499}\inst{4},
        Michele Cappellari\orcidicon{0000-0002-1283-8420}\inst{5},
        Shude Mao\orcidicon{0000-0001-8317-2788}\inst{4}
      }
\institute{ National Astronomical Observatories, Chinese Academy of Sciences, 20A Datun Road, Chaoyang District, Beijing 100101, China; {\it ranl@bao.ac.cn}\\
         \and
         Institute for Frontiers in Astronomy and Astrophysics, Beijing Normal University, Beijing 102206, China\\
         \and
         School of Astronomy and Space Science, University of Chinese Academy of Sciences, Beijing 10049, China\\
         \and
         Department of Astronomy, Tsinghua University, Beijing 100084, China\\
         \and
         Sub-department of Astrophysics, Department of Physics, University of Oxford, Denys Wilkinson Building, Keble Road, Oxford, OX1 3RH, UK\\
 \vs \no
   {\small Received 2023 April 8; accepted 2023 April 26}
}

\abstract{
The second moment of the stellar velocity within the effective radius, denoted by $\sigma_{\rm e}^2$, is a crucial quantity in galaxy studies as it provides insight into galaxy properties and their mass distributions. However, large spectroscopic surveys typically do not measure $\sigma_{\rm e}$ directly, instead providing $\sigma_{\rm aper}$, the second moment of the stellar velocity within a fixed fiber aperture. In this paper, we derive an empirical aperture correction formula, given by $\sigma_{\rm aper}/\sigma_{\rm e}=(R_{\rm aper}/R_{\rm e})^{\alpha}$, using spatially resolved stellar kinematics extracted from approximately 10,000 Sloan Digital Sky Survey–Mapping Nearby Galaxies at Apache Point Observatory (SDSS-MaNGA) integral field unit observations. Our analysis reveals a strong dependence of $\alpha$ on the $r$-band absolute magnitude $M_{\rm r}$, $g-i$ color, and Sersic index $n_{\rm Ser}$, where $\alpha$ values are lower for brighter, redder galaxies with higher Sersic indices. Our results demonstrate that the aperture correction derived from previous literature on early-type galaxies cannot be applied to predict the aperture corrections for galaxies with intermediate Sersic indices. We provide a lookup table of $\alpha$ values for different galaxy types, with parameters in the ranges of $-18>M_{\rm r}>-24$, $0.4<g-i<1.6$, and $0<n_{\rm Ser}<8$. A Python script is provided to obtain the correction factors from the lookup table.
\keywords{galaxies:  evolution  –  galaxies:  formation  –  galaxies:  kinematics and dynamics – galaxies: structure}}

\authorrunning{Zhu et al.}            
   \titlerunning{Velocity dispersion aperture corrections in MaNGA}  
   \maketitle
\section{Introduction} 
\label{sec:intro}
Stellar kinematics provide crucial information for understanding the mass distributions of galaxies. The advent of integral field spectroscopy (IFS) surveys, such as SAURON \citep{deZeeuw2002}, $\rm ATLAS^{3D}$ \citep{Cappellari2011}, CALIFA \citep{Sanchez2012}, SAMI \citep{Bryant2015}, and MaNGA \citep{Bundy2015}, has allowed for spatially resolved stellar kinematics, which, when combined with well-established dynamical modelling methods, such as Schwarzschild modelling \citep{Schwarzschild1979} and Jeans Anisotropic Modelling \citep[JAM;][]{Cappellari2008,Cappellari2020}, can provide accurate measurements of the mass distribution of galaxies \citep{Cappellari2006,Cappellari2013a,Scott2015,ZhuL2018NatAs,Li2018a,DynPop1}. The scaling relations between the dynamical properties  and the stellar populations of the galaxies can be explored in detail \citep[e.g.][]{Cappellari2016ARA&A,Li2018a,DynPop2}.

However, the high cost of obtaining spatially resolved kinematics limits the applicability of IFS to a significantly larger number of local galaxies, such as the millions of galaxies in the Sloan Digital Sky Survey (SDSS), or distant galaxies at high redshift that cannot be spatially resolved. Thus, the second moment of stellar velocity within an aperture $\sigma_{\rm aper}$, which is measured using single-fiber spectroscopy, remains a fundamental quantity for understanding the dynamics of galaxies. Important scaling relations such as the Fundamental Plane \citep{Dressler1987,Djorgovski1987,Jorgensen1995} and the Mass Plane \citep[e.g.,][]{Auger2010,Cappellari2013a,DynPop3} are derived with respect to $\sigma_{\rm aper}$  as well. However, galaxies have different angular sizes spanning a wide dynamic range, but the fiber size is fixed. As a result, velocity moments are not measured coherently for the entire galaxy sample. Therefore, researchers need to correct the measured velocity dispersion to a physically meaningful radius, which is typically the effective radius of a galaxy \citep{Auger2010,Chen2019,deGraaff2021}.

Earlier studies on velocity dispersion aperture corrections focused on early-type galaxies, with velocity dispersion profiles typically described as a power-law function of the form shown in \autoref{eq:vd_profile}.  \citet{Jorgensen1995} derived a power-law slope of $\alpha=-0.04$ for the early-type galaxies in nine clusters, while  \citet{Mehlert2003} found $\alpha=-0.063$ for 35 early-type galaxies in the Coma cluster. A later study based on the SAURON IFS data analyzed 40 early-type galaxies and reported a slope of $\alpha=-0.066\pm0.035$  \citep{Cappellari2006}. The aperture corrections were extended to 300 CALIFA galaxies across the Hubble sequence  \citep{Falcon-Barroso2017}, which found a consistent power-law slope for the early-type galaxies ($\alpha=-0.055\pm0.020$) and a strong variation of $\alpha$ with magnitude (or stellar mass) for the late-type galaxies ($\alpha=0.047\pm0.021$ for $M_{\rm r}<-22$, $\alpha=0.086\pm0.013$ for $-22<M_{\rm r}<-20$, $\alpha=0.153\pm0.063$ for $M_{\rm r}>-20$). Despite the $\alpha$ variation with magnitude, the non-negligible scatter in a given $M_{\rm r}$ bin (especially for the late-type galaxies of $M_{\rm r}>-20$) indicates a potential secondary contributor to the variation of $\alpha$. The unprecedently large sample of the MaNGA survey provides the ability to perform a further and more comprehensive analysis of the velocity dispersion aperture correction. Recently,  \citet{deGraaff2021} derived $\alpha=-0.033\pm0.003$ from the MaNGA data but they only selected a subset of MaNGA sample (702 galaxies).

In this work, we take advantage of the full sample ($\sim10000$) of MaNGA IFS observations to perform a more detailed classification of galaxies and try to obtain more accurate aperture corrections for each type of galaxy. We investigate the relations between the shape of velocity dispersion profiles (quantified by a power-law form) and other properties, e.g. magnitude, color, and Sersic index \citep{Sersic1968}. By selecting galaxies within a narrow parameter range, we aim to eliminate the effect of sample bias and provide a lookup table that can be applied to various types of galaxies, resulting in more precise aperture corrections.

The organization of this paper is as follows. In \autoref{sec:data}, we provide a brief overview of the stellar kinematic data and the MaNGA sample. Our main results are presented in \autoref{sec:results}. Finally, we summarize our conclusions in \autoref{sec:conclusions}. We adopt a standard cosmology with $\Omega_{\rm m} = 0.3$, $\Omega_{\rm \Lambda}=0.7$, and $H_0 = 70\,\mathrm{km\,s^{-1}\,Mpc^{-1}}$ throughout this work.

\section{Data}
\label{sec:data}
\subsection{The MaNGA survey}
As a part of the SDSS-IV, the MaNGA \citep{Bundy2015} is an integral field unit (IFU) survey that uses tightly packed arrays of optical fibers to obtain spectral measurements of approximately 10,000 nearby galaxies. Using the BOSS spectrographs  \citep{Smee2013,Drory2015} on the Sloan 2.5 m telescope  \citep{Gunn2006} at the Apache Point Observatory, this survey covers a radial range up to 1.5 effective radii ($R_{\rm e}$) for the Primary+ sample and up to 2.5 $R_{\rm e}$ for the Secondary sample  \citep{Law2015, Wake2017}. MaNGA provides spatially resolved spectra with a spaxel size of $0.5\arcsec$ and the average $g-$band Point Spread Function (PSF) FWHM throughout the survey is about $2.54\arcsec$  \citep{Law2016}.

The spectral measurements across the wavelength range of $\rm 3600-10300\,\AA$ have a spectral resolution of $\sigma = 72\,\rm km\,s^{-1}$  \citep{Law2016}. The raw observational data require spectrophotometric calibration \citep{Yan2016}, which is performed using the Data Reduction Pipeline (DRP)  \citep{Law2016}. The DRP processes the data to produce three-dimensional data cubes that can be used to create spatially resolved maps of the galaxies under observation.

\subsection{Stellar kinematics}
The Data Analysis Pipeline (DAP;  \citealt{Belfiore2019,Westfall2019}) is responsible for producing higher-level products such as stellar kinematics, nebular emission-line properties, and spectral indices of galaxies. The stellar kinematic information is derived from the IFU spectra using the \textsc{ppxf} software  \citep{Cappellari2004,Cappellari2017,Cappellari2022}, which fits absorption lines with a subset of the MILES  \citep{Sanchez-Blazquez2006,Falcon-Barroso2011} stellar library, MILES-HC. The spectra are Voronoi binned  \citep{Cappellari2003} to $\rm S/N=10$ to ensure the reliability of the derived stellar velocity dispersions, which are presented as a combination of the intrinsic velocity dispersion of stars ($\sigma_{\ast}$) and the quadrature difference between the instrumental dispersion of the galaxy template and the MaNGA data ($\sigma_{\rm diff}$). The velocity dispersion of the galaxy can be obtained using the equation $\sigma_{\ast}^2=\sigma_{\rm obs}^2-\sigma_{\rm diff}^2$, where $\sigma_{\rm obs}$ is the observed velocity dispersion \citep{Westfall2019}. In this work, we make use of the maps of stellar velocity, stellar velocity dispersion, and $g-$band flux, which are taken from the DAP outputs\footnote{\url{https://www.sdss.org/dr17/manga/manga-data/data-access/}}.

\subsection{Sample selection}
We obtain 10,735 DAP outputs from SDSS DR17  \citep{Abdurro'uf2022}, which includes 10,296 observations of galaxies and the ancillary program targets such as the Coma, IC342, M31, and globular clusters. After excluding 151 flagged datacubes that have been identified as critical-quality or unusual-quality from 10,296 galaxy observations, there are 10,145 high-quality datacubes corresponding to 10,010 unique galaxies and 135 repeat observations. The sample has a nearly uniform distribution of stellar masses in the range of $10^9-6\times10^{11}\,\rm M_{\odot}$ and a median redshift of approximately 0.03 \citep{Wake2017}.

\begin{figure*}
    \centering
    \includegraphics[width=1.7\columnwidth]{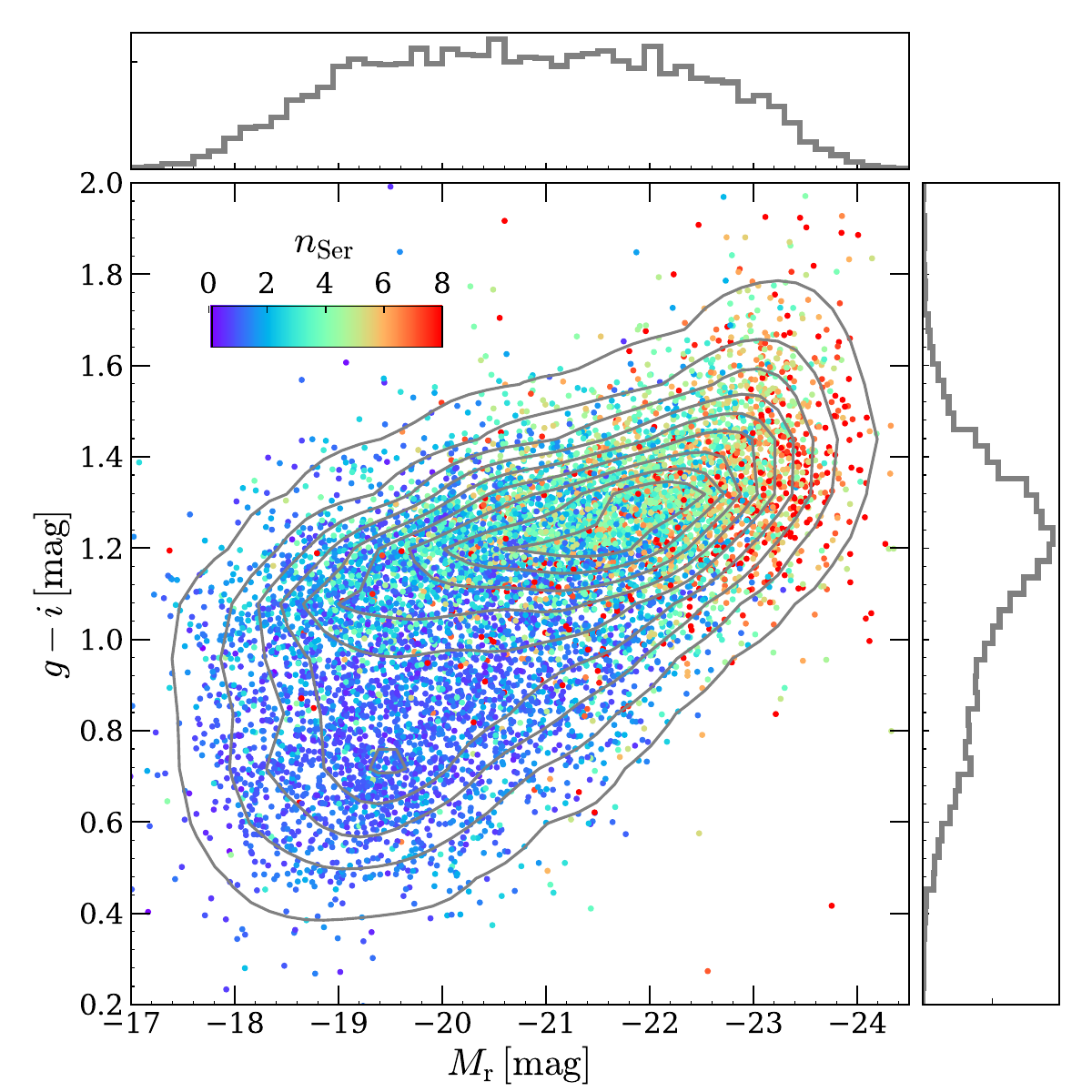}
    \caption{Color-magnitude diagram, with symbols color-coded by the Sersic index. All the quantities ($r$-band absolute magnitude, $g-i$ color, and Sersic index $n_{\rm Ser}$) are taken from the PyMorph catalog  \citep{pymorph2022}. The gray contours are the two-dimensional number density distributions, while the gray histograms on the top and right are the one-dimensional distributions (normalized to unity).}
    \label{fig:color_mag_nser}
\end{figure*}

In this work, we use the stellar kinematics of 10,010 unique galaxies but remove the galaxies that have been identified as having bad stellar kinematics \citep{DynPop1}. Finally, we obtain a sample of 9132 galaxies. The distribution of the whole sample is presented in \autoref{fig:color_mag_nser}, in which the color $g-i$, the $r-$band absolute magnitude $M_{\rm r}$, and the Sersic index are taken from the PyMorph catalog  \citep{pymorph2022}. As shown in \autoref{fig:color_mag_nser}, the sample spans a wide range of galactic properties: from the faint ($M_{\rm r}=-18$ mag) to the bright ($M_{\rm r}=-24$ mag), and from the blue ($g-i=0.4$ mag) to the red ($g-i=1.6$ mag). The large sample and various types of galaxies (i.e. the red sequence, the blue cloud, and the green valley) enable us to comprehensively study the aperture effects on the velocity dispersion.

\section{Results}
\label{sec:results}
\subsection{Integrated Stellar Velocity Dispersion Profiles}
We derive the integrated stellar velocity dispersion within a set of apertures and obtain the aperture profiles for the whole sample. The integrated velocity dispersion is defined as the flux-weighted second velocity moments within a given elliptical aperture of area $A = \pi R_{\rm aper}^2$, written as
\begin{equation}
\label{eq:sigma_aper}
    \sigma_{\rm aper}\approx\langle v_{\rm rms}^2\rangle_{\rm aper}^{1/2}=\sqrt{\frac{\sum_k F_k (V_k^2+\sigma_k^2)}{\sum_k F_k}},
\end{equation}
where $V_k$, $\sigma_k$, and $F_k$ are the stellar velocity, stellar velocity dispersion, and $g-$band flux in the $k$th spaxel, respectively. Following  \citet[section~3.3.3]{Cappellari2013a}, we choose to use the elliptical aperture instead of the circular aperture to properly account for the inclination effects. Using elliptical apertures is also more appropriate for flat galaxies, which are typically dominated by rotation. If a circular aperture defined by the effective radius is used, the peak of rotation may be located outside the aperture, resulting in an underestimated $\sigma_{\rm aper}$. Since the $V$ and $\sigma$ of DAP outputs are derived from the Voronoi binned spectra, we assign the binned values to each $0.5\arcsec\times0.5\arcsec$ spaxel belonging to each Voronoi bin. We calculate $\sigma_{\rm aper}$ within the elliptical apertures with fixed ellipticity and position angle (PA), which are derived from the single-component Sersic fits  \citep{pymorph2022}. 

\begin{figure*}
    \centering
    \includegraphics[width=\textwidth]{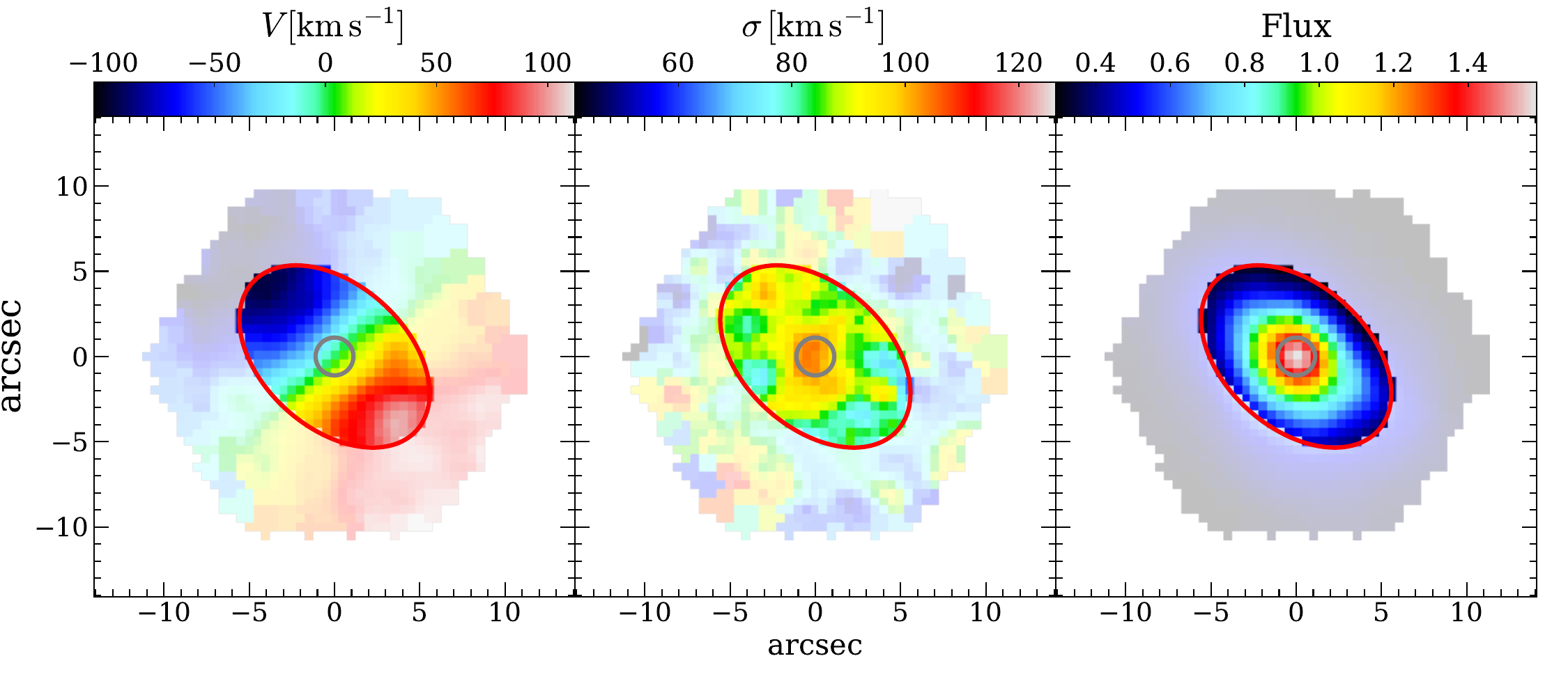}
    \caption{Illustration for the definition of $\sigma_{\rm aper}$ (MaNGA ID: 12-180432). The panels from left to right are the maps of stellar velocity, stellar velocity dispersion, and $g-$band flux. The gray circle denotes the $\rm FWHM_{\rm PSF}/2.355$, while the red ellipse is the half-light isophote of area $A=\pi R_{\rm e}^2$, where $R_{\rm e}$ is the effective radius. The $R_{\rm e}$, ellipticity, and position angle are taken from the single Sersic fits  \citep{pymorph2022}. The $\sigma_{\rm e}$ is calculated using \autoref{eq:sigma_aper} and the pixels within the red ellipse, while $\sigma_{\rm aper}$ values are determined within the concentric ellipses of area $A=\pi R_{\rm aper}^2$.}
    \label{fig:stellar_kinematics_example}
\end{figure*}
\begin{figure}
    \centering
    \includegraphics[width=\columnwidth]{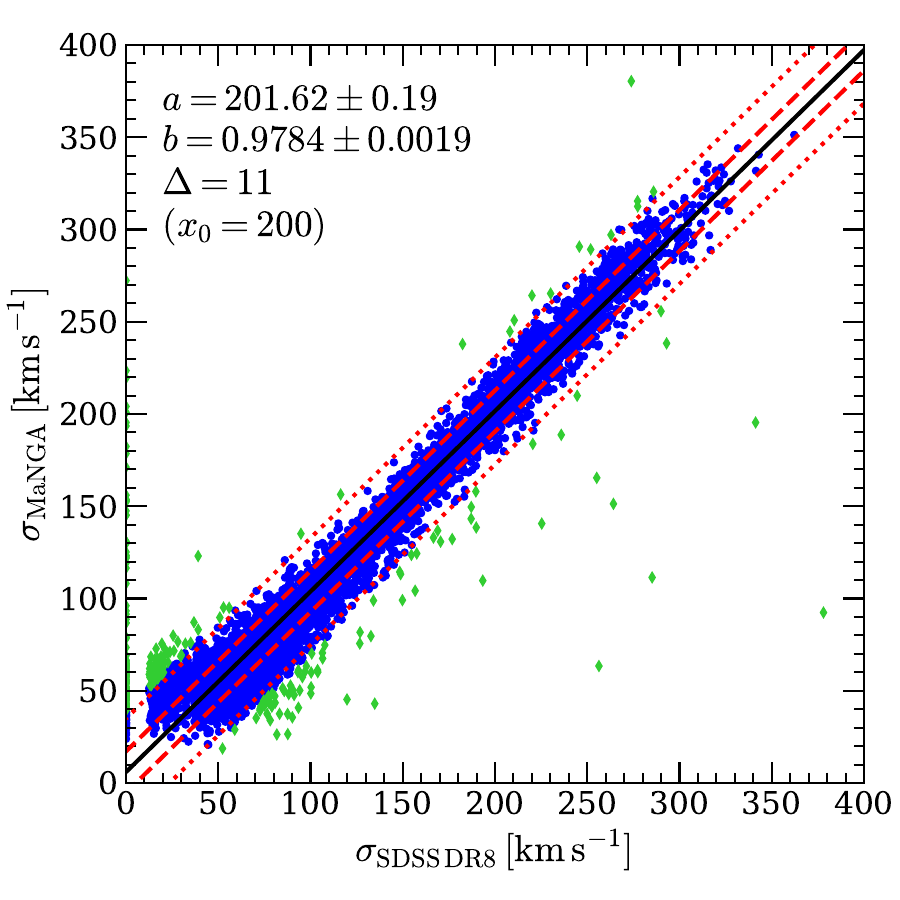}
    \caption{Comparison of the stellar velocity dispersion derived from the MaNGA survey with the SDSS DR8 measurements  \citep{SDSSDR8} within the same circular aperture of $3\arcsec$ diameter. The velocity dispersions of MaNGA are computed using \autoref{eq:sigma_aper}, while those from SDSS DR8 are from the single fit of the integrated spectra with the same aperture. The black solid, red dashed, and red dotted lines are the best-fitting, 1$\sigma$, and 2.6$\sigma$ lines obtained using the \textsc{lts\_linefit} procedure (with \texttt{clip=3}). The coefficients of the best-fitting $y=a+b\times(x-x_{0})$ are shown in the panel, while $\Delta$ is the observed rms scatter. The green symbols are the detected outliers of \textsc{lts\_linefit} beyond 3$\sigma$ confidence level.}
    \label{fig:cmp_MaNGA_SDSS}
\end{figure}

In \autoref{fig:stellar_kinematics_example}, we present an example to illustrate the calculation of integrated stellar velocity dispersion within a given elliptical aperture. In this figure, the pixels within the red ellipse of area $A=\pi R_{\rm e}^2$ are adopted to estimate the effective velocity dispersion $\sigma_{\rm e}$ using \autoref{eq:sigma_aper}. The effective velocity dispersion $\sigma_{\rm e}$ is demonstrated to agree well with that measured from a single fit on the stacked spectra within the same aperture  \citep{Cappellari2013a}. Moreover, we also calculate the integrated velocity dispersion within a circular aperture with a diameter of 3$\arcsec$ like the SDSS single fibers using \autoref{eq:sigma_aper}. Then we match the galaxies with the spectroscopic catalog of SDSS DR8\footnote{\url{https://www.sdss3.org/dr8/data_access.php}} \citep{SDSSDR8} and compare the velocity dispersion values obtained with two methods in \autoref{fig:cmp_MaNGA_SDSS}. To ensure that the velocity dispersions derived from SDSS single-fiber spectra are reliable, we only select the spectral measurements with median $\rm S/N>10$. In \autoref{fig:cmp_MaNGA_SDSS}, we perform a linear fit to the two quantities with the \textsc{lts\_linefit}\footnote{\url{https://pypi.org/project/ltsfit/}} software  \citep{Cappellari2013a}, which combines the Least Trimmed Squares robust technique of  \citet{Rousseeuw2006} into a least-squares fitting algorithm and allows for the intrinsic scatter and errors in all coordinates. As shown in the figure, the slope ($b=0.9784\pm0.0019$) and the small rms scatter ($\Delta=11\,\rm km\,s^{-1}$) denote the high consistency between the two measurements, while most detected outliers are potential unreliable measurements with velocity dispersions below the resolution limit ($\rm 100\,km\,s^{-1}$) of SDSS spectrograph. This justifies the accuracy of \autoref{eq:sigma_aper} in estimating the integrated stellar velocity dispersion within a given aperture from IFU kinematics.

\begin{figure*}
    \centering
    \includegraphics[width=\textwidth]{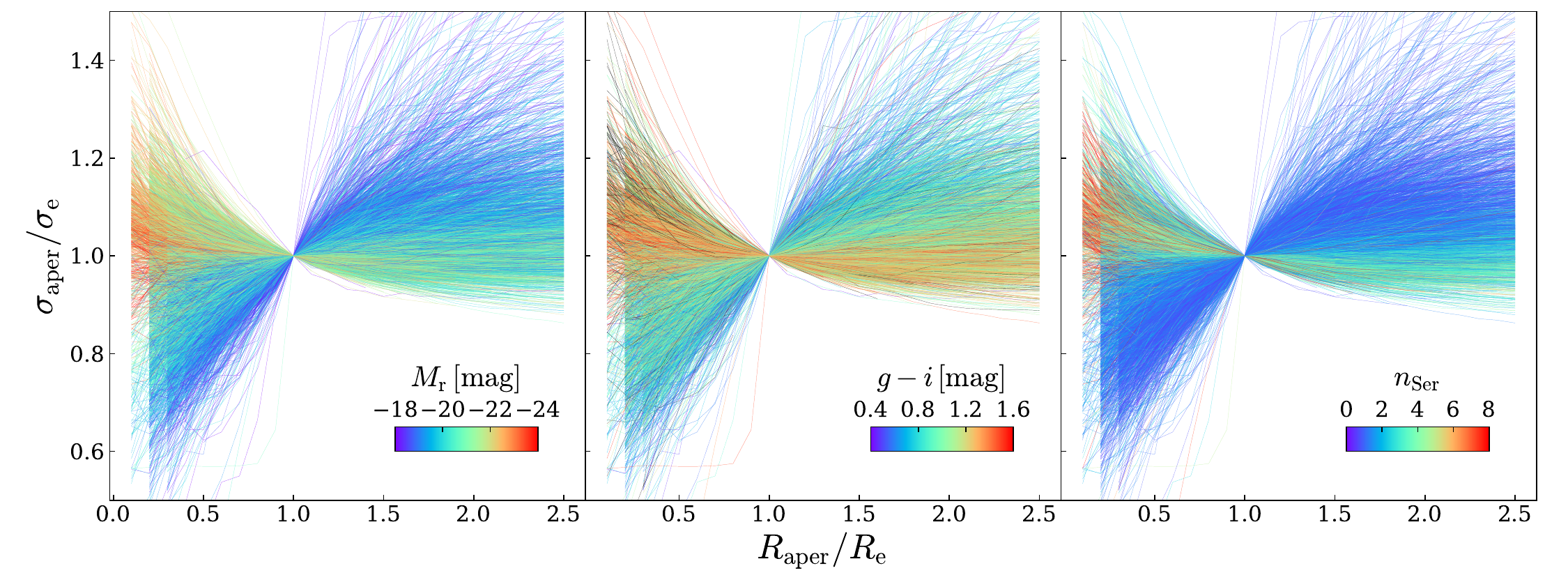}
    \caption{The stellar velocity dispersion integrated within elliptical apertures (the area equals $\pi R_{\rm aper}^2$) as a function of radius. The profiles are normalized by the effective radius $R_{\rm e}$ and the dispersion $\sigma_{\rm e}$ within an elliptical aperture of area $A=\pi R_{\rm e}$. The profiles are colored by $r-$band absolute magnitude $M_{\rm r}$ (left), $g-i$ color (middle), and Sersic index $n_{\rm Ser}$ (right).}
    \label{fig:vd_profiles}
\end{figure*}

For each galaxy, we calculate the $\sigma_{\rm aper}$ within a set of elliptical apertures of area $A=\pi R_{\rm aper}^2$. $R_{\rm aper}$ ranges from 0.1$R_{\rm e}$ to 2.5$R_{\rm e}$, with a linear step of 0.1$R_{\rm e}$. We use the $R_{\rm e}$ and the effective velocity dispersion $\sigma_{\rm e}$ as normalization factors to rescale the integrated velocity dispersion profiles. The normalized profiles are presented in \autoref{fig:vd_profiles}, which are colored by the SDSS $r-$band  \citep{Stoughton2002} absolute magnitude $M_{\rm r}$ (left panel), the color $g-i$ (middle panel), and the Sersic index $n_{\rm Ser}$ (right panel). As can be seen, the $\sigma_{\rm aper}$ profiles vary significantly with different galactic properties: the brighter and redder galaxies with higher Sersic index tend to have decreasing $\sigma_{\rm aper}$ profiles, while the fainter and bluer galaxies with lower Sersic index show increasing trends toward outside. The increasing (decreasing) trends of $\sigma_{\rm aper}$ profiles are due to their different $V/\sigma$ profiles. The $\sigma_{\rm aper}$ of massive and red galaxies with higher Sersic index tend to be dispersion-dominated and the dispersions decrease with increasing radius, while the less massive and blue galaxies are rotation-supported and the $\sigma_{\rm aper}$ profiles increase in tandem with rotation curves.

The dependencies of $\sigma_{\rm aper}$ profiles on other galactic properties are consistent with the large variations of $\sigma_{\rm aper}$ profiles across a wide range of morphological types observed in \citet{Falcon-Barroso2017}, which analyzed the $\sigma_{\rm aper}$ profiles for 300 CALIFA galaxies \citep{Sanchez2012}. However, the significantly larger sample (approximately 10,000 galaxies) of MaNGA enables a more detailed analysis and provides more accurate aperture corrections for integrated velocity dispersion measurements for specific types of galaxies.

\subsection{Slopes of $\sigma_{\rm aper}$ Profiles for Different Types of Galaxies}

\begin{figure}
    \centering
    \includegraphics[width=\columnwidth]{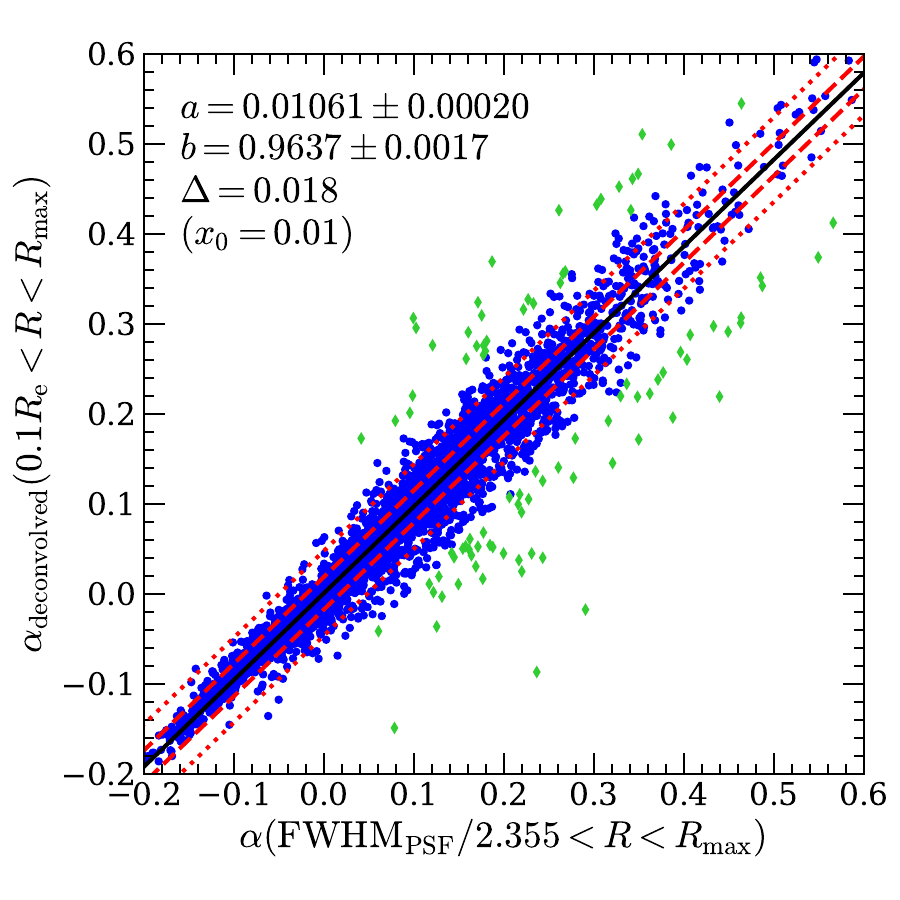}
    \caption{Comparison of the power-law index $\alpha$ from fitting within ${\rm FWHM}_{\rm PSF}/2.355<R<R_{\rm max}$ (\autoref{eq:fitting_range}) with those deconvolved values fitting within $0.1R_{\rm e}<R<R_{\rm max}$. The deconvolved $\alpha$ values are obtained from convolving the MaNGA PSF on the model $\sigma_{\rm aper}$ profiles during the fitting process. The lines and symbols are the same as \autoref{fig:cmp_MaNGA_SDSS} but with a \texttt{clip=5} in the \textsc{lts\_linefit} fitting, which means that the green symbols are the outliers beyond 5$\sigma$ confidence level.}
    \label{fig:cmp_observe_deconvolve}
\end{figure}

Following previous works in the literature  \citep{Jorgensen1995,Cappellari2006,Falcon-Barroso2017}, we use a power-law function to quantify the slopes of the normalized $\sigma_{\rm aper}$ profiles as
\begin{equation}
\label{eq:vd_profile}
    \left(\frac{\sigma_{\rm aper}}{\sigma_{\rm e}}\right)=\left(\frac{R_{\rm aper}}{R_{\rm e}}\right)^{\alpha}.
\end{equation}
We fit the individual $\sigma_{\rm aper}$ profiles for each galaxy within a range of [$R_{\rm in}$, $R_{\rm out}$] defined as
\begin{subequations}
\label{eq:fitting_range}
\begin{align}
    R_{\rm in}={\rm max}({\rm FWHM}/2.355, 0.1R_{\rm e}),\\
    R_{\rm out}={\rm min}(R_{\rm max}, 2.5R_{\rm e}),
\end{align}
\end{subequations}
where the FWHM is in $g-$band (from DAP) and $R_{\rm max}$ is the largest radius of Voronoi bins, to avoid the PSF effects on the determination of $\alpha$ parameters. As discussed in section 6 of  \citet{Falcon-Barroso2017}, beam smearing may affect the integrated velocity dispersion at the very center of galaxies and lead to potential bias in measuring $\alpha$ parameters. However, we argue here that $\alpha$ will not be clearly affected because most data points of the $\sigma_{\rm aper}$ profiles are the integrated velocity dispersion within much larger apertures compared to the dispersion of PSF. Following  \citet{Falcon-Barroso2017}, we tried to convolve the MaNGA PSF to the model $\sigma_{\rm aper}$ profiles when fitting them to the observed ones. In \autoref{fig:cmp_observe_deconvolve}, we present the comparison between the $\alpha$ with PSF deconvolved (fitting in a range of [0.1$R_{\rm e}$, $R_{\rm max}$]) and the $\alpha$ with PSF un-deconvolved (fitting in a range of [FWHM/2.355, $R_{\rm max}$]), and find the two values are in good agreement with a slope of $0.9637\pm0.0017$ and a small rms scatter of $\Delta=0.018$ using the \textsc{lts\_linefit} procedure. Practically, aperture correction is usually applied to correct the velocity second moment measured within a fiber aperture that is comparable to or even larger than the effective radius of the galaxy, to the effective radius. The PSF smearing effect at the inner region of the MaNGA data is not expected to introduce a significant bias to this correction. In our project, we did not include the PSF effect in our fiducial analysis.

\begin{figure*}
    \centering
    \includegraphics[width=\textwidth]{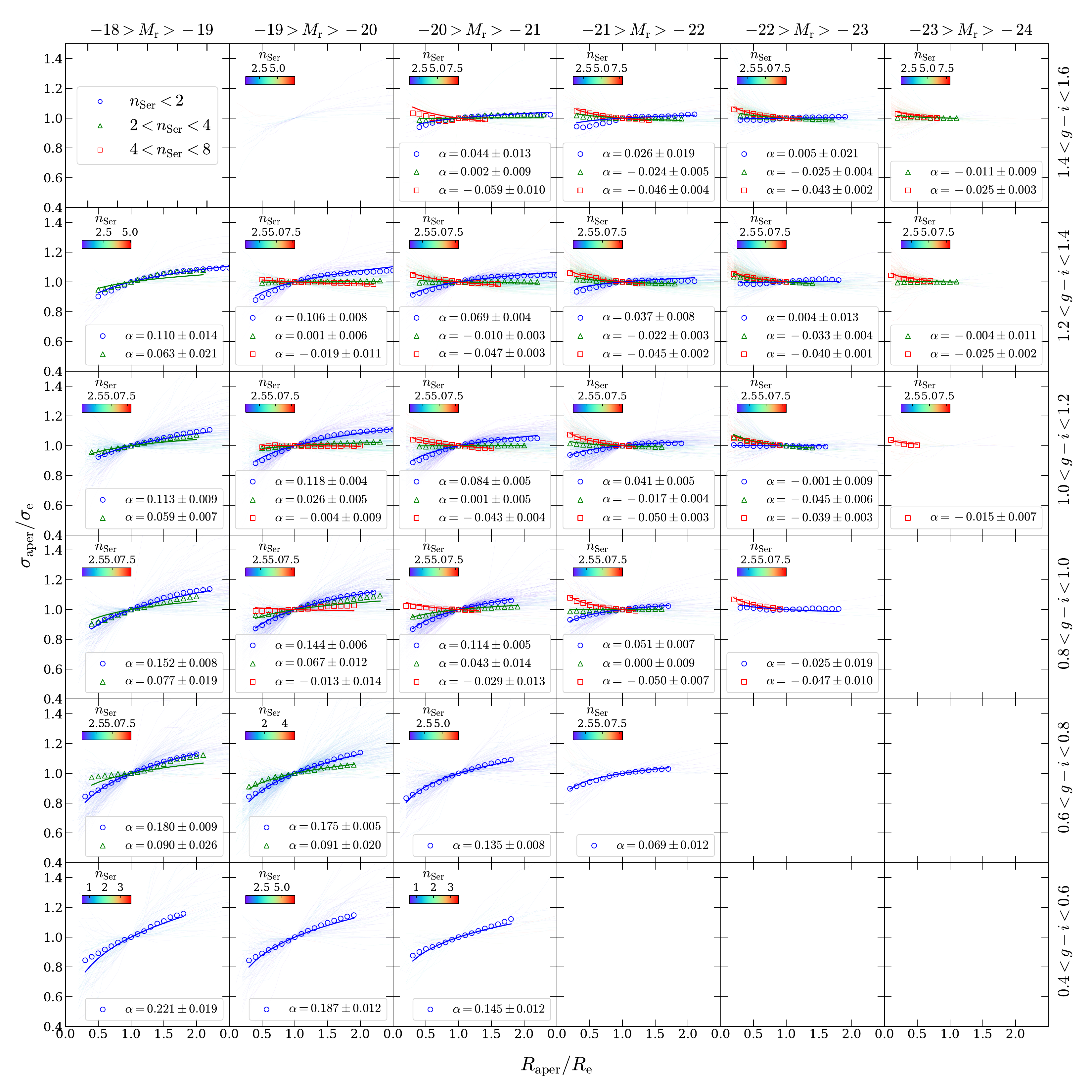}
    \caption{Normalized integrated velocity dispersion profiles in different $M_{\rm r}$ and $g-i$ bins, where $M_{\rm r}$ is the $r-$band absolute magnitude and $g-i$ color is the difference between $g-$band and $i-$band absolute magnitudes. The $\sigma_{\rm aper}$ profiles are colored by the Sersic index $n_{\rm Ser}$. In each panel within given ($M_{\rm aper}$, $g-i$) bin, the biweight mean profiles for different $n_{\rm Ser}$ are shown as blue circles ($n_{\rm Ser}<2$), green triangles ($2<n_{\rm Ser}<4$), and red squares ($4<n_{\rm Ser}<8$). The mean and standard deviation of $\alpha$ (see the text at the bottom of panels) for each subset of galaxies (at least 15) are obtained from bootstrapping with 100 iterations. The curves (blue, green, red) overlaid on the symbols are the corresponding modelled $\sigma_{\rm aper}$ profiles, which are described in the form of a power-law function (\autoref{eq:vd_profile}).}
    \label{fig:vdslope_color_mag}
\end{figure*}

We divide the full sample into different subsamples based on their $r-$band absolute magnitude $M_{\rm r}$ and $g-i$ color. There are 6 $M_{\rm r}$ bins ranging from $-$18 to $-$24 (from faint to bright) in a step of 1 mag and 6 bins of $g-i$ color from 0.4 to 1.6 (from blue to red). In \autoref{fig:vdslope_color_mag}, we present the $\sigma_{\rm aper}$ profiles color-coded by Sersic index, which still show large variations in a given narrow bin of ($M_{\rm r}$, $g-i$). Thus, in each ($M_{\rm r}$, $g-i$) bin, we further split the galaxies into different Sersic index groups, i.e. $n_{\rm Ser}<2$, $2<n_{\rm Ser}<4$, $4<n_{\rm Ser}<8$. For each subsample ($M_{\rm r}$, $g-i$, $n_{\rm Ser}$) with at least 15 galaxies, we adopt the biweight  \citep{Hoaglin1983} mean value of $\alpha$. The mean and standard deviation of $\alpha$ parameters (presented in each panel of \autoref{fig:vdslope_color_mag}) are obtained from bootstrapping with 100 iterations. In each panel, the biweight mean $\sigma_{\rm aper}$ profiles with different Sersic indices (denoted as blue circles, green triangles, and red squares) can be described well with the predicted power-law relations (solid lines) using the mean $\alpha$ values.

We create a lookup table\footnote{A Python script is provided in \url{https://github.com/kaizhu-astro/aperture_correction} to obtain the correction factors from the lookup table.} (\autoref{tab:alpha}) of aperture corrections for various types of galaxies by assuming a power-law function of $\sigma_{\rm aper}$ profiles. The table lists the mean and standard deviation of $\alpha$, which are determined from bootstrapping with 100 iterations, for galaxies within given ($M_{\rm r}$, $g-i$, $n_{\rm Ser}$) bins.

\begin{figure*}
    \centering
    \includegraphics[width=\textwidth]{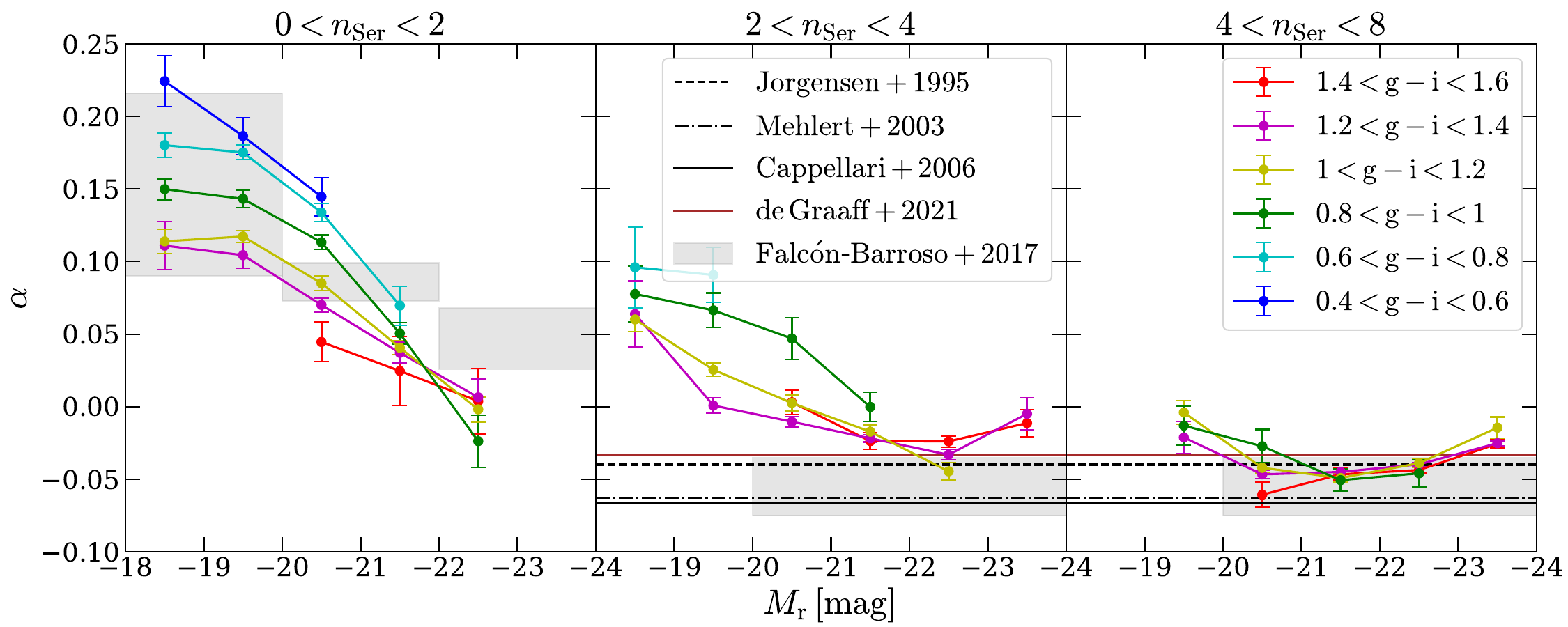}
    \caption{Power-law index $\alpha$ of integrated velocity dispersion profiles as a function of $r-$band absolute magnitude $M_{\rm r}$ for different $g-i$ color (see the legends in the right panel) and Sersic index $n_{\rm Ser}$ (see the top of each panel). The horizontal black dashed line is $\alpha=-0.04$ from  \citet{Jorgensen1995}, the horizontal black dashed-dotted line is $\alpha=-0.063$ from  \citet{Mehlert2003}, and the black solid line is $\alpha=-0.066$ from the SAURON project  \citep{Cappellari2006}. The value of MaNGA subset (brown horizontal line), $\alpha=-0.033$, is taken from  \citet{deGraaff2021}, whose sample is dominated by the high Sersic index galaxies but is not selected based on morphology. The gray-shaded squares are taken from the CALIFA survey \citep{Falcon-Barroso2017}, which derived $\alpha$ for both early-type and late-type galaxies.}
    \label{fig:cmp_vdslope}
\end{figure*}

In \autoref{fig:cmp_vdslope}, we investigate the relations between $\alpha$ parameters and $M_{\rm r}$ for galaxies with different $g-i$ color and Sersic index. Overall, the $\alpha$ values are smaller for brighter (smaller $M_{\rm r}$), redder (higher $g-i$) galaxies with higher $n_{\rm Ser}$. The smaller $\alpha$ parameters (i.e., the galaxies have larger central $\sigma_{\rm aper}$) for galaxies with higher $n_{\rm Ser}$ are due to the fact that a larger $n_{\rm Ser}$ implies a greater concentration of central stellar mass, resulting in an increased central $\sigma_{\rm aper}$ and a steeper $\sigma_{\rm aper}$ profile. 

The left panel of \autoref{fig:cmp_vdslope} shows that for galaxies with $0<n_{\rm Ser}<2$, which are typically classified as late-type galaxies, strong decreasing trends of $\alpha$ with increasing luminosity and redder color are observed. This suggests the importance of detailed classification when applying aperture corrections to late-type galaxies. The trend is consistent with the fact that redder spiral galaxies tend to host large bulges, resulting in increased central densities and larger central $\sigma_{\rm aper}$ values than their blue counterparts. It is worth noting that the variation in $g-i$ color is responsible for the scatter of $\alpha$ in the $-18>M_{\rm r}>-20$ bin, as compared to the $\alpha$ values of CALIFA late-type galaxies \citep{Falcon-Barroso2017}. In the brightest bin ($-22>M_{\rm r}>-24$), the $\alpha$ values of MaNGA late-type galaxies tend to be slightly smaller than those in \citet{Falcon-Barroso2017}, likely due to the latter deriving values from galaxies with positive $\alpha$ instead of selecting galaxies based on Sersic index or morphology.

Similar but weaker trends, where brighter and redder galaxies have smaller $\alpha$, are also observed in the sample with $2<n_{\rm Ser}<4$. However, for galaxies with $4<n_{\rm Ser}<8$, they tend to have negative $\alpha$ values. In this bin, the $\alpha$-$M_r$ relation shows a flat U-shape, and the scatter among different color bins is small.

Our results for the $4<n_{\rm Ser}<8$ bin are consistent with previous studies on early-type galaxies. For example, \citet{Jorgensen1995} found $\alpha=-0.04$, \citet{Mehlert2003} found $\alpha=-0.063$, SAURON \citep[][section 2.3]{Cappellari2006} found $\alpha=-0.066\pm0.035$, CALIFA \citep[][section 6]{Falcon-Barroso2017} found $\alpha=-0.055\pm0.020$, and a subset of MaNGA galaxies \citep[][appendix C]{deGraaff2021} found $\alpha=-0.033\pm0.003$. It should be noted that the subset of the MaNGA sample used in \citet{deGraaff2021} is dominated by galaxies with high Sersic index but is not a pure early-type galaxy sample, so the $\alpha$ value for early-type galaxies is slightly shallower than what we found.

The results of this paper demonstrate that the aperture correction derived from previous literature on early-type galaxies cannot be applied to predict the aperture corrections for galaxies with intermediate Sersic indices, regardless of their color.

\begin{table*}
\centering
\caption{A lookup table for the aperture correction factors of integrated stellar velocity dispersion. The galaxies are divided into 6 bins in $r-$band absolute magnitude $M_{\rm r}$ (for different columns), 6 bins in $g-i$ color (for different rows), and 3 bins in Sersic index $n_{\rm Ser}$ (for different sub rows). In total, there are 108 ($M_{\rm r}$, $g-i$, $n_{\rm Ser}$) bins. In each bin with a sample of at least 15 galaxies, the mean and standard deviation of $\alpha$ (see the definition in \autoref{eq:vd_profile}) are estimated from bootstrapping with 100 iterations. One also needs to account for the effect of aperture shape when applying the correction factors on the real observations with circular apertures (see discussions in \autoref{sec:circular_correction}).}
\label{tab:alpha}
\scriptsize
\begin{tabular}{|c|c|c|c|c|c|c|c|c|}
\hline
\multicolumn{2}{|c}{\multirow{2}*{}}&\multicolumn{6}{|c|}{$M_{\rm r}$}&\multicolumn{1}{|c|}{\multirow{2}*{$n_{\rm Ser}$}}\\
\cline{3-8}
\multicolumn{1}{|c}{}& &[-19, -18] & [-20, -19] & [-21, -20] & [-22, -21] & [-23, -22] & [-24, -23]&\multicolumn{1}{c|}{}\\
\hline
\multicolumn{1}{|c|}{\multirow{18}*{$g-i$}} & \multirow{3}*{[1.4, 1.6]} &  - & - &$0.044\pm0.013$ & $0.026\pm0.019$&$0.005\pm0.021$& -&[0, 2] \\
\cline{3-9}
&&-&-&$0.002\pm0.009$&$-0.024\pm0.005$&$-0.025\pm0.004$&$-0.011\pm0.009$&[2, 4]\\
\cline{3-9}
&&-&-&$-0.059\pm0.010$&$-0.046\pm0.004$&$-0.043\pm0.002$&$-0.025\pm0.003$&[4, 8]\\
\cline{2-9}
 & \multirow{3}*{[1.2, 1.4]} & $0.110\pm0.014$&$0.106\pm0.008$ &$0.069\pm0.004$ &$0.037\pm0.008$ &$0.004\pm0.013$ &- &[0, 2] \\
 \cline{3-9}
 &&$0.063\pm0.021$&$0.001\pm0.006$&$-0.010\pm0.003$&$-0.022\pm0.003$&$-0.033\pm0.004$&$-0.004\pm0.011$&[2, 4] \\
 \cline{3-9}
 &&-&$-0.019\pm0.011$&$-0.047\pm0.003$&$-0.045\pm0.002$&$-0.040\pm0.001$&$-0.025\pm0.002$&[4, 8]\\
\cline{2-9}
\cline{2-9}
 & \multirow{3}*{[1.0, 1.2]} &$0.113\pm0.009$ &$0.118\pm0.004$ &$0.084\pm0.005$ &$0.041\pm0.005$ &$-0.001\pm0.009$ &- &[0, 2] \\
 \cline{3-9}
&&$0.059\pm0.007$&$0.026\pm0.005$&$0.001\pm0.005$&$-0.017\pm0.004$&$-0.045\pm0.006$&-&[2, 4] \\
\cline{3-9}
&&-&$-0.004\pm0.009$&$-0.043\pm0.004$&$-0.050\pm0.003$&$-0.039\pm 0.003$&$-0.015\pm0.007$&[4, 8]\\
\cline{2-9}
 & \multirow{3}*{[0.8, 1.0]} &$0.152\pm0.008$ &$0.144\pm0.006$ &$0.114\pm0.005$ &$0.051\pm0.007$ &$-0.025\pm0.019$ &- &[0, 2] \\
 \cline{3-9}
&&$0.077\pm0.019$&$0.067\pm0.012$&$0.043\pm0.014$&$0.000\pm0.009$&-&-&[2, 4]\\
\cline{3-9}
&&-&$-0.013\pm0.014$&$-0.029\pm0.013$&$-0.050\pm0.007$&$-0.047\pm0.010$&-&[4, 8]\\
\cline{2-9}
 & \multirow{3}*{[0.6, 0.8]} & $0.180\pm0.009$&$0.175\pm0.005$ &$0.135\pm0.008$ &$0.069\pm0.012$ &- &- &[0, 2] \\
 \cline{3-9}
&&$0.090\pm0.026$&$0.091\pm0.020$&-&-&-&-&[2, 4]\\
\cline{3-9}
&&-&-&-&-&-&-&[4, 8]\\
\cline{2-9}
 & \multirow{3}*{[0.4, 0.6]} & $0.221\pm0.019$& $0.187\pm0.012$ &$0.145\pm0.012$ &- &- &- &[0, 2] \\
 \cline{3-9}
&&-&-&-&-&-&-&[2, 4]\\
\cline{3-9}
&&-&-&-&-&-&-&[4, 8]\\
\hline
\end{tabular}
\end{table*}

\subsection{Corrections for the Velocity Dispersion Measured within a Circular Aperture}
\label{sec:circular_correction}
The clear trends that $\alpha$ varies with $M_{\rm r}$, $g-i$, and $n_{\rm Ser}$ highlight the enhanced accuracy of aperture corrections in this study. However, it is difficult to apply such aperture corrections in reality due to the fact of circular apertures in single-fiber observations. To further account for the effect of aperture shape, we also calculate the $\sigma_{\rm aper, circ}$, which is also defined as \autoref{eq:sigma_aper} but within a circular aperture with a radius of $R_{\rm aper}$. We fit the $\sigma_{\rm aper, circ}$ profiles for each galaxy, which are normalized as
\begin{equation}
\label{eq:vd_cir_profile}
    \frac{\sigma_{\rm aper, circ}}{\sigma_{\rm e}}=\left(\frac{R_{\rm aper}}{R_{\rm e}}\right)^{\alpha_{\rm circ}},
\end{equation}
to obtain the power-law index $\alpha_{\rm circ}$. The $\alpha_{\rm circ}$ can be applied in real single-fiber observations to obtain the $\sigma_{\rm e}$.

\begin{figure}
    \centering
    \includegraphics[width=\columnwidth]{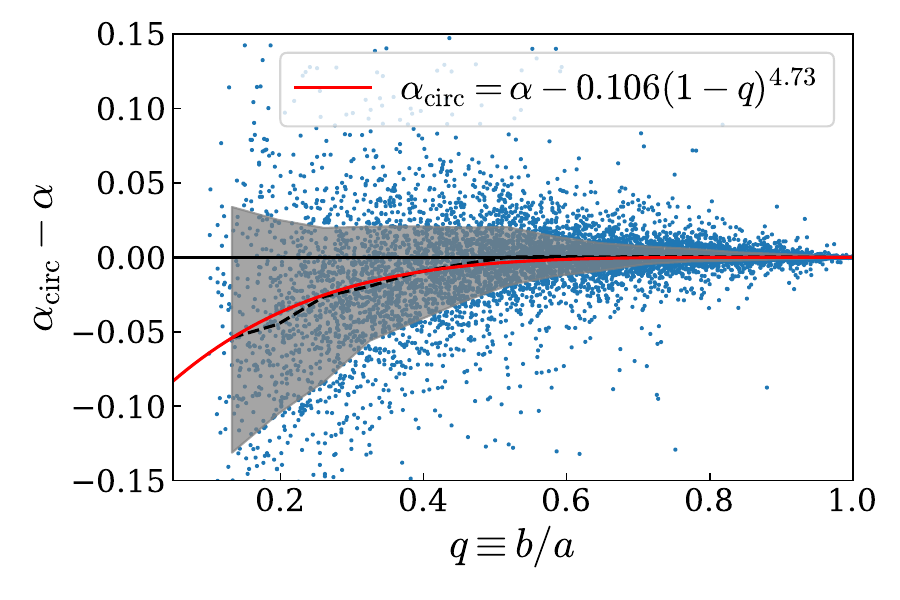}
    \caption{The $\alpha_{\rm circ}-\alpha$ as a function of axial ratio $q$. The $\alpha$ and $\alpha_{\rm circ}$ are derived from \autoref{eq:vd_profile} and \autoref{eq:vd_cir_profile} respectively, while the $q$ is derived from Sersic fits  \citep{pymorph2022}. The black dashed curve and the gray-shaded region are the median value and [16th, 84th] percentiles. The red curve is the best-fitting empirical relation for all galaxies.}
    \label{fig:alphadiff_axialratio}
\end{figure}

In \autoref{fig:alphadiff_axialratio}, we show the $\alpha_{\rm circ}-\alpha$ as a function of axial ratio $q\equiv b/a$. As can be seen, the systematic difference between $\alpha$ and $\alpha_{\rm circ}$ is negligible for galaxies with $q>0.4$, while flat galaxies ($q<0.4$) tend to have smaller $\alpha_{\rm circ}$ than $\alpha$. Given the small fraction ($\sim 20\%$) of galaxies with $q<0.4$ in our sample, we do not present the $\alpha_{\rm circ}$ values in each ($M_{\rm r}$, $g-i$, $n_{\rm Ser}$), which are expected to be similar to $\alpha$. As shown in \aref{appendix:test}, similar trends can be seen if replacing $\alpha$ in \autoref{fig:cmp_vdslope} with $\alpha_{\rm circ}$. However, we also found that the biweight mean $\sigma_{\rm aper, circ}/\sigma_{\rm e}$ profiles in some ($M_{\rm r}$, $g-i$, $n_{\rm Ser}$) bins cannot be well described by the function, especially for the $n_{\rm Ser}<2$ galaxies that may suffer from strong inclination effects. This is likely due to the fact that the normalization of $\sigma_{\rm aper, circ}/\sigma_{\rm e}$ is nonphysical and will bring uncertainties when stacking the normalized profiles in a given ($M_{\rm r}$, $g-i$, $n_{\rm Ser}$) bin. Thus, we choose to only present the $\alpha$ values and derive an empirical relation between $\alpha$ and $\alpha_{\rm circ}$, which relates to the axial ratio $q$ in the form of
\begin{equation}
\label{eq:ell2cir}
    \alpha_{\rm circ} = \alpha-0.106(1-q)^{4.73}.
\end{equation}
One can use \autoref{tab:alpha} to obtain $\alpha$ in a given ($M_{\rm r}$, $g-i$, $n_{\rm Ser}$) bin and then use \autoref{eq:ell2cir} to obtain the $\alpha_{\rm circ}$ if necessary ($q<0.4$), but one should also be aware of the large scatter of ($\alpha_{\rm circ}-\alpha$) for very flat galaxies.

\section{Conclusions}
\label{sec:conclusions}

We conduct a comprehensive analysis of the aperture corrections for the integrated stellar velocity dispersion $\sigma_{\rm aper}$ using the full MaNGA sample. With a large sample size of approximately 10,000 galaxies, we are able to study aperture corrections in detail for  a diverse range of galaxy types (\autoref{fig:color_mag_nser}) for the first time. We derived the $\sigma_{\rm aper}$ profiles for the entire sample (\autoref{fig:vd_profiles}) and used a power-law function (\autoref{eq:vd_profile}) to quantify the profiles of different subsamples based on their $r-$band absolute magnitude $M_{\rm r}$, $g-i$ color, and Sersic index $n_{\rm Ser}$ (\autoref{fig:vdslope_color_mag}). The relationships between the power-law index $\alpha$ and the three properties ($M_{\rm r}$, $g-i$, $n_{\rm Ser}$) are presented in \autoref{tab:alpha} and \autoref{fig:cmp_vdslope}.

Our analysis reveals several important findings regarding the aperture corrections of the integrated stellar velocity dispersion $\sigma_{\rm aper}$. Firstly, we observe decreasing trends of the power-law index $\alpha$ with increasing $M_{\rm r}$, redder $g-i$ color, and higher Sersic index $n_{\rm Ser}$. While the $\alpha$ values of early-type galaxies with high $n_{\rm Ser}$ ($4<n_{\rm Ser}<8$) are consistent with previous studies in the literature, those of early-type galaxies with $2<n_{\rm Ser}<4$ cannot be predicted by models of previous studies. Secondly, we find that for late-type galaxies ($n_{\rm Ser}<2$), the median $\alpha$ varies significantly with $M_{\rm r}$ and $g-i$ color, with the largest variation in $\alpha$ on $g-i$ color being up to $\sim 0.1$ in $-19<M_{\rm r}<-18$. These diverse $\alpha$ values and their clear correlations with other galactic properties highlight the importance of a comprehensive analysis based on the large sample of MaNGA.

In addition, we establish an empirical relation between the power-law index $\alpha$ and $\alpha_{\rm circ}$, as shown in \autoref{eq:ell2cir}. Here, $\alpha_{\rm circ}$ is defined as the power-law index derived from the velocity dispersion profile measured within circular apertures, which can be used to calculate $\sigma_{\rm e}$ for circular apertures with any given radius defined by the size of observational fiber. We believe that our findings will enable more precise aperture corrections for single-fiber spectroscopic survey, such as those from the SDSS and Dark Energy Spectroscopic Instrument survey (DESI). However, our empirical relations are derived from the MaNGA sample with a median redshift of $z=0.03$ and some care should be taken when extrapolated to higher redshift galaxies. A possible solution could be using the higher redshift IFS observations (e.g. Multi Unit Spectroscopic Explorer; MUSE) to calibrate the relations.

\begin{acknowledgements}
We acknowledge the support of National Nature Science Foundation of China (Nos 11988101,12022306), the National Key R$\&$D Program of China No. 2022YFF0503403,  the support from the Ministry of Science and Technology of China (Nos. 2020SKA0110100),  the science research grants from the China Manned Space Project (Nos. CMS-CSST-2021-B01,CMS-CSST-2021-A01), CAS Project for Young Scientists in Basic Research (No. YSBR-062), and the support from K.C.Wong Education Foundation.

Funding for the Sloan Digital Sky 
Survey IV has been provided by the 
Alfred P. Sloan Foundation, the U.S. 
Department of Energy Office of 
Science, and the Participating 
Institutions. 

SDSS-IV acknowledges support and 
resources from the Center for High 
Performance Computing  at the 
University of Utah. The SDSS 
website is www.sdss.org.

SDSS-IV is managed by the 
Astrophysical Research Consortium 
for the Participating Institutions 
of the SDSS Collaboration including 
the Brazilian Participation Group, 
the Carnegie Institution for Science, 
Carnegie Mellon University, Center for 
Astrophysics | Harvard \& 
Smithsonian, the Chilean Participation 
Group, the French Participation Group, 
Instituto de Astrof\'isica de 
Canarias, The Johns Hopkins 
University, Kavli Institute for the 
Physics and Mathematics of the 
Universe (IPMU) / University of 
Tokyo, the Korean Participation Group, 
Lawrence Berkeley National Laboratory, 
Leibniz Institut f\"ur Astrophysik 
Potsdam (AIP),  Max-Planck-Institut 
f\"ur Astronomie (MPIA Heidelberg), 
Max-Planck-Institut f\"ur 
Astrophysik (MPA Garching), 
Max-Planck-Institut f\"ur 
Extraterrestrische Physik (MPE), 
National Astronomical Observatories of 
China, New Mexico State University, 
New York University, University of 
Notre Dame, Observat\'ario 
Nacional / MCTI, The Ohio State 
University, Pennsylvania State 
University, Shanghai 
Astronomical Observatory, United 
Kingdom Participation Group, 
Universidad Nacional Aut\'onoma 
de M\'exico, University of Arizona, 
University of Colorado Boulder, 
University of Oxford, University of 
Portsmouth, University of Utah, 
University of Virginia, University 
of Washington, University of 
Wisconsin, Vanderbilt University, 
and Yale University.
\end{acknowledgements}

\appendix
\section{Tests on using $\alpha_{\rm circ}$ to predict $\sigma_{\rm aper, circ}$ profiles}
\label{appendix:test}
\begin{figure*}
    \includegraphics[width=\textwidth]{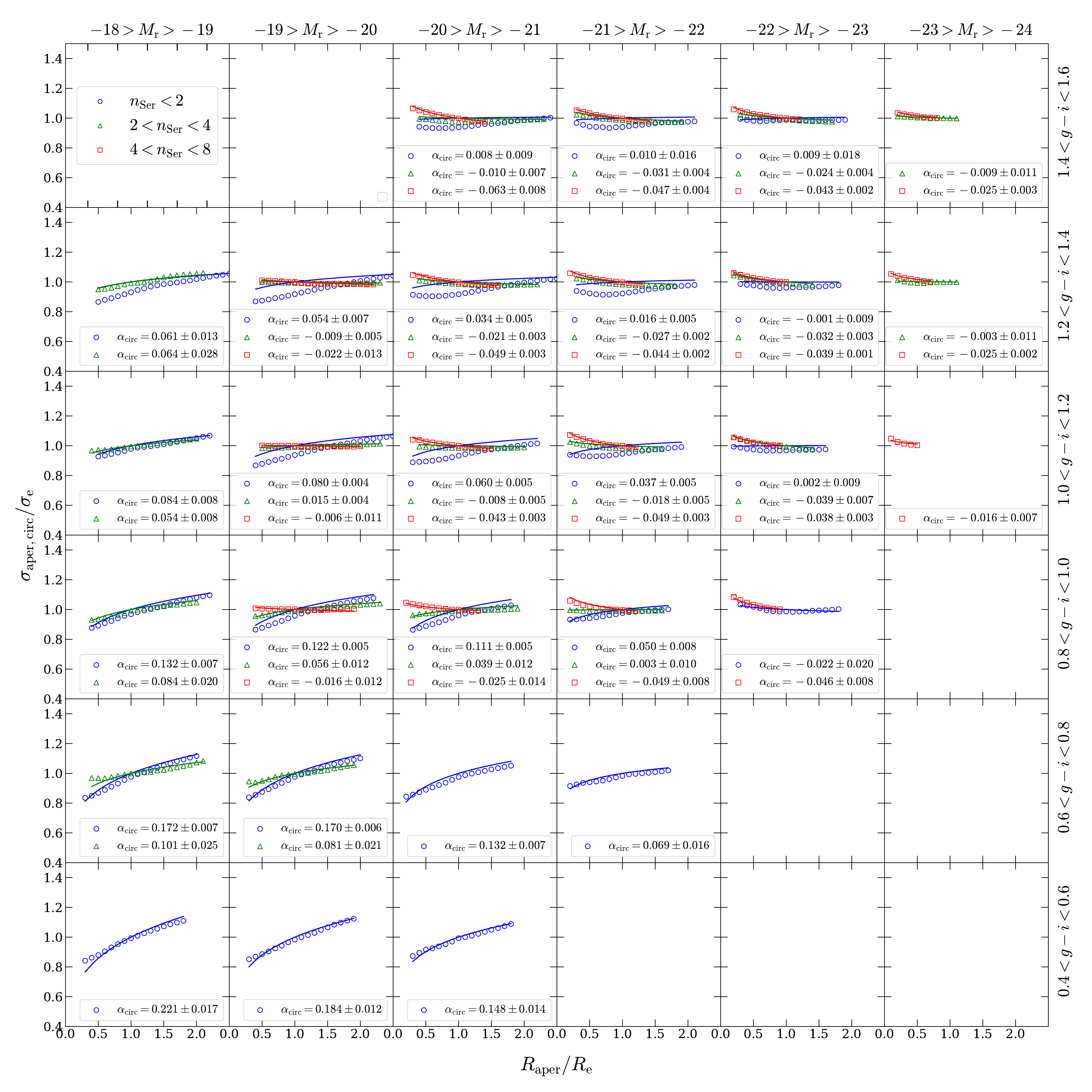}
    \caption{The same as \autoref{fig:vdslope_color_mag}, but replacing $\sigma_{\rm aper}$ with $\sigma_{\rm aper, circ}$.}\label{fig:vdslope_circ_color_mag}
\end{figure*}

\begin{figure*}
    \includegraphics[width=\textwidth]{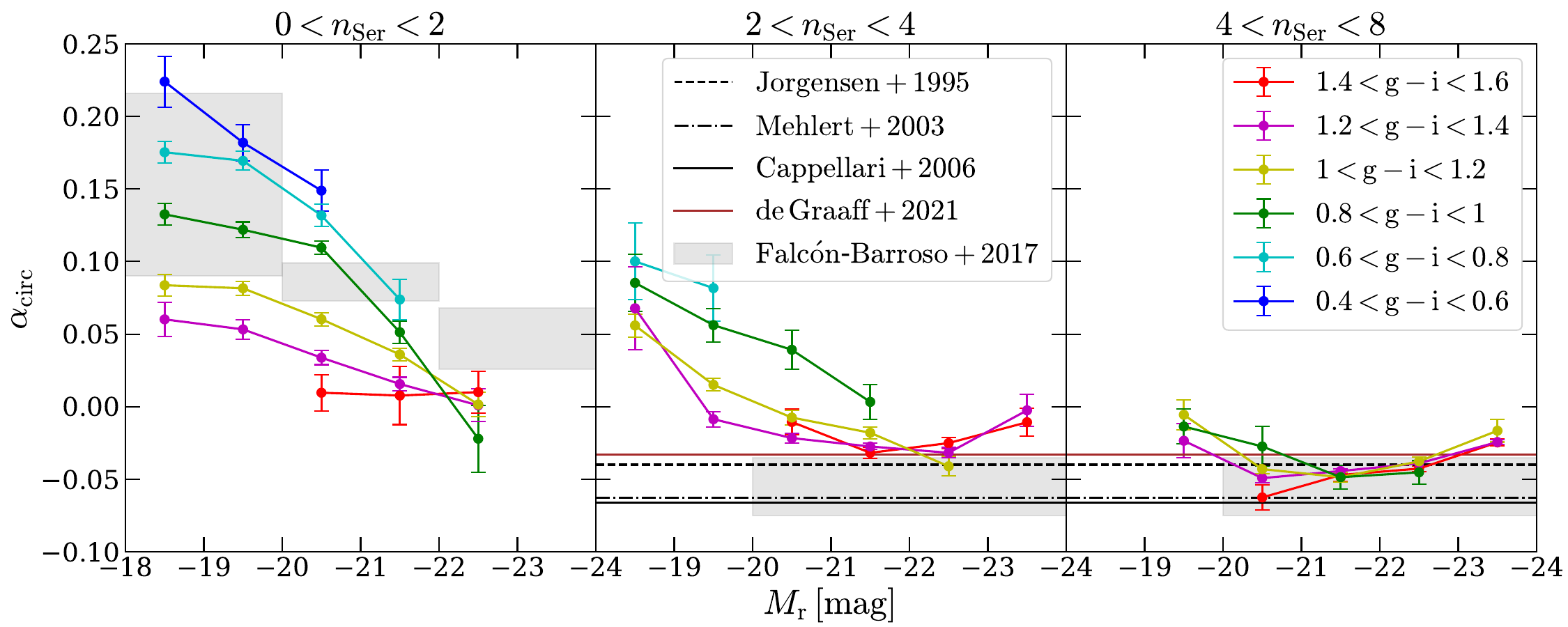}
    \caption{The same as \autoref{fig:cmp_vdslope}, but replacing $\alpha$ with $\alpha_{\rm circ}$.}\label{fig:cmp_vdslope_circ}
\end{figure*}

We present two figures (\autoref{fig:vdslope_circ_color_mag} and \autoref{fig:cmp_vdslope_circ}) that are similar to \autoref{fig:vdslope_color_mag} and \autoref{fig:cmp_vdslope}, but $\sigma_{\rm aper}$ profiles and $\alpha$ are replaced with $\sigma_{\rm aper, circ}$ profiles and $\alpha_{\rm circ}$, respectively. We do not recommend directly using $\alpha_{\rm circ}$ to predict $\sigma_{\rm aper, circ}$ profiles. See the text in \autoref{sec:circular_correction} for a detailed discussion.

\bibliography{ref}
\bibliographystyle{raa}

\end{document}